\documentclass[twocolumn,noshowpacs,pre,preprintnumbers,superscriptaddress,amsmath,amssymb,floatfix]{revtex4}
\usepackage[caption=false]{subfig}
\usepackage{graphicx}
\usepackage{color}
\usepackage{placeins}
\usepackage[normalem]{ulem}

\newcommand{\ktot}{\langle k_{total} \rangle}

\newcommand{\kintra}{\langle k_{intra} \rangle}
\newcommand{\kh}{\langle k_H \rangle}
\newcommand{\kv}{\langle k_V \rangle}

\newcommand{\mkkij}{\langle k_{i,j} \rangle}

\begin{document}
	
	\title{Cascading failures in anisotropic interdependent networks of spatial modular structures}
	\author{Dana Vaknin}
	\affiliation{Department of Physics, Bar-Ilan University, Ramat Gan, Israel}
	\author{Amir Bashan}
	\affiliation{Department of Physics, Bar-Ilan University, Ramat Gan, Israel}
	\author{Lidia A. Braunstein}
	\affiliation{Instituto de Investigaciones Físicas de Mar del Plata (IFIMAR)-Departamento de Física, FCEyN, Universidad Nacional de Mar del Plata-CONICET, Déan Funes 3350, (7600) Mar del Plata, Argentina}
	\author{Sergey V. Buldyrev}
	\affiliation{Department of Physics, Yeshiva University, New York, USA}
	\author{Shlomo Havlin}
	\affiliation{Department of Physics, Bar-Ilan University, Ramat Gan, Israel}
	
	\date{\today}
	
	\begin{abstract}
		
		The structure of real-world multilayer infrastructure systems usually exhibits anisotropy due to constraints of the embedding space. For example, geographical features like mountains, rivers and shores influence the architecture of critical infrastructure networks. Moreover, such spatial networks are often non-homogeneous but rather have a modular structure with dense connections within communities and sparse connections between neighboring communities. When the networks of the different layers are interdependent, local failures and attacks may propagate throughout the system. Here we study the robustness of spatial interdependent networks which are both anisotropic and heterogeneous. We also evaluate the effect of localized attacks having different geometrical shapes. We find that anisotropic networks are more robust against localized attacks and that anisotropic attacks, surprisingly, even on isotropic structures, are more effective than isotropic attacks.
		
	\end{abstract}
	\maketitle
	
	\section{Introduction}
    
    Many real-world systems are well correlated with a population density that is highly non-uniform and concentrates in large cities or is spread along seacoasts, rivers, or major transportation routes. 
    Such systems are influenced by geographical and social features and usually combine both spaciality on large scales and randomness on small scales.
    More specifically, in many cases such as infrastructure networks, the connections within the cities are dense and almost uniform, while the connections between cities are mainly between nearby cities.
    These factors were the motivation behind the profound studies of spatial networks \cite{girvan-pnas2002,guimera-pnas2005,palla-nature2005,kosmidis-epl2008,bradde-prl2010,mucha-science2010,barthelemy-physicsreports2011,bashan-naturephysics2013,du-chaos2014,danziger-epl2016}, either homogeneous or heterogeneous and composed of communities.
    However, the connections between the cities do not have to be isotropic. For example, if in one axis there are more topographic obstacles (e.g. mountains or rivers) then in that axis there will be fewer connections than in the other axes.
    Yet, most models of spatial networks are isotropic, i.e, have no preferred orientation in the network structure.
    
    In addition, many realistic systems such as modern infrastructure networks can be considered as strongly interdependent multi-layered networks \cite{rinaldi-ieee2001,chang-bridge2009,hines-proceedings2010}.
    Therefore, great consideration was given to the study of interdependent networks and in particular their robustness, using percolation theory \cite{buldyrev-nature2010,parshani-prl2010,gao-pre2012,baxter-prl2012,dedomenico-prx2013,son-epl2012,dong-pre2014,kivela-jcomnets2014,boccaletti-physicsreports2014,radicchi-naturephysics2015,shekhtman-njp2015,bianconi-oxfordjournals2018}.
    Such studies have been carried out to analyze cascading failures resulting from various attacks \cite{motter-pre2002,wei-prl2012,gao-naturephysics2012,reis-naturephysics2014,yuan-pre2015,vaknin-njp2017,spiewak-netwsci2018}. One of the key results in these types of studies is that often localized attacks are significantly more effective when compare to random attacks.
    To the best of our knowledge, in studies of localized attacks, the attack itself has always been uniform within a certain radius \cite{berezin-scireports2015,shao-njp2015,yang-science2017,vaknin-prr2020} although in the real world this ``uniformity condition" usually does not occur. For example, natural disasters (such as an earthquake) and malicious targeted attacks can be anisotropic and even contained within a single axis.
    Here, we take into consideration {\bf anisotropy} in modeling both in the systems and in the localized attacks.

	\begin{figure}
		\centering
		\includegraphics[width=0.95\linewidth]{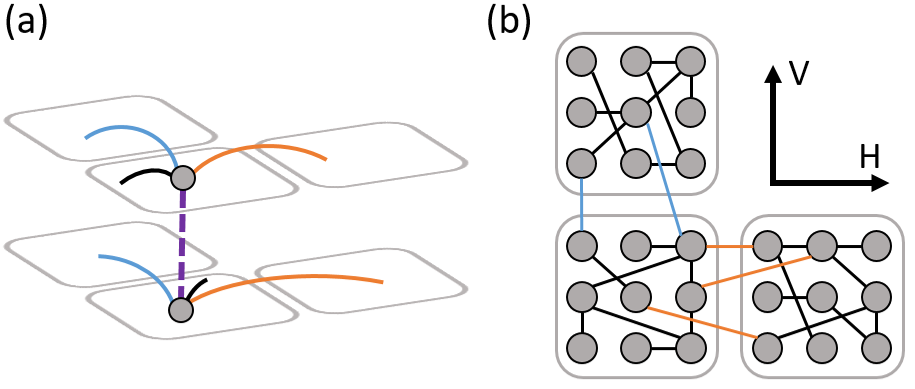}
		\caption{
			\textbf{Anisotropy in a model of interdependent spatial networks.}
			\textbf{(a)} Schematic representation of neighboring communities (gray squares) in a multiplex model of two layers. 
			The nodes are represented by gray circles. 
			The connectivity links can be divided into three types, as follows: ‘intra-links’ which connect nodes within the communities (black links), ‘horizontal-links’ (orange) and ‘vertical-links’ (blue) which connect pairs of nodes from neighboring communities in different orientations.
			Nodes from the two layers are interdependent (dashed purple link). 
			\textbf{(b)} The system is constructed as a $m\times m$ lattice of communities, each of which is an ER network of nodes located at the sites of the $\zeta\times\zeta$ square lattice. For clarity here we show $3$ ER networks of size $\zeta=3$.
			In the simulations we set periodic boundary conditions and we solve the limit of infinitely large ER communities, i.e., the case of $\zeta \rightarrow \infty$.
			Anisotropy is modeled by different degree distributions of horizontal-links and vertical-links (three orange links and two blue links). 
		}
		\label{fig:Demonstration2}
	\end{figure}
	
	\section{Model}
	
    Here we introduce an anisotropic model of interdependent spatially embedded networks with communities. 
    The model assumes that the entire 2D territory is divided into squared communities of size $\zeta\times\zeta$ representing cities or densely populated areas, see Fig. \ref{fig:Demonstration2}.
    For simplicity, we assume a multiplex system with two topologically distinct network layers that share the same set of nodes.
    The interdependence between the layers is introduced as follows: if a node becomes dysfunctional in one layer it is also dysfunctional in the other layer.
    In each layer, the connectivity links within a community (‘intra-links’) are connected at random, i.e. forming ER networks, while the links connecting nodes in two different communities (‘inter-links’) can only connect neighboring squares in each network, horizontally or vertically, as illustrated in Fig. \ref{fig:Demonstration2}(b).
    Each node has a degree $k_{intra}$ of intra-links, a degree $k_H$ of horizontal-links, a degree $k_V$ of vertical-links, and the total degree is $k_{total}=k_{intra}+k_H+k_V$. We assume that $k_{intra}$, $k_H$ and $k_V$ are independent random variables taken from three different degree distributions which are characterized by average degrees $\kintra$, $\kh$ and $\kv$, respectively. 
    The anisotropy of the system is specified by the parameter
    \begin{equation}
    \label{eq:gamma}
    \gamma = \frac{\kh}{\kh+\kv},
    \end{equation}
    which is the ratio between the degree of the horizontal-links and the inter-links (and thus $\gamma=1/2$ is the isotropic case), and the heterogeneity of the system is specified by the ratio between the degree of the inter-link and the total degree $\alpha = (\kh+\kv)/\ktot$.
	
	Here we study the robustness of spatial anisotropic multiplex networks to different forms of localized attacks.
	Initial damage in a multiplex network spreads in a process in which failures of some nodes lead to failures of other nodes and so on.
	In the cascading failures process, a node fails if it is no longer connected to the giant component in one of the layers.
	Thus, the cascading stops when all the remaining nodes are part of the mutual giant component (MGC), i.e., all the remaining nodes are connected to the giant component in both layers.
	We consider a network as functioning if the MGC is of size $O(N)$, where $N$ is the number of nodes in the network.
	Accordingly, in our spatial multilayer network model, the network is considered to be resistant to a localized attack if the damage spreads to a finite number of communities and does not reach the edges of the system.
	
	\section{Analytical Approach}
	
	In our recent study \cite{vaknin-prr2020}, we developed a mathematical framework that consists of general equations for the MGC size of a multiplex model of $m \times m$ ER communities connected as a lattice. 
    However, in that study, we applied the equations only for the case of an isotropic network and we analyzed only the effect of isotropic localized attacks on the functionality of the network.
	Here, we study a more general and realistic case, of an anisotropic model with different average degrees for the vertical links $\kv$ and the horizontal links $\kh$, as well as anisotropic localized attacks.
    For a network with a given levels of heterogeneity and anisotropy, we determine whether it remains functional after the cascading failure induced by isotropic or anisotropic initial damages. 
	The initial damage in community $i$ (for $i$ from 1 to $m^2$) is expressed by the parameter $p_i$, which is defined as the fraction of nodes that survived as a result of the damage.
	In the analytical model we consider infinitely large ER communities ($\zeta \rightarrow \infty$) that are characterized by Poisson degree distribution.
	Using the analytical formalism from our recent study \cite{vaknin-prr2020}, we obtain the following equation for the MGC size of each community $i$, $P_{\infty,i}$,
	\begin{equation}
	\label{eq:Pinf}
        P_{\infty,i} = p_i \cdot [1-e^{\sum_j\langle k_{i,j}\rangle P_{\infty,j}}]^2,
    \end{equation}
    where
    \begin{equation}
    \label{eq:kij}
        \langle k_{i,j}\rangle = \left\{\begin{matrix}
        \langle k_{intra} \rangle  & \textup{for }  i=j
        \\ 
        \langle k_H \rangle/2  & \textup{if $j$ is horizontal neighbor of $i$} 
        \\
        \langle k_{V} \rangle/2  & \textup{if $j$ is vertical neighbor of $i$}  
        \\
        0  & \textup{else}  .
        \end{matrix}\right.
    \end{equation}
    
    Next we use this analytical formalism (Eq. (\ref{eq:Pinf})) to evaluate the robustness of interdependent spatial systems characterized by heterogeneity and anisotropy by calculating the steady state of each community.
    In the Appendix, we verify these analytical solutions through simulations.
    
	\section{Results}
	
	\begin{figure}
		\centering
		\includegraphics[width=1\linewidth]{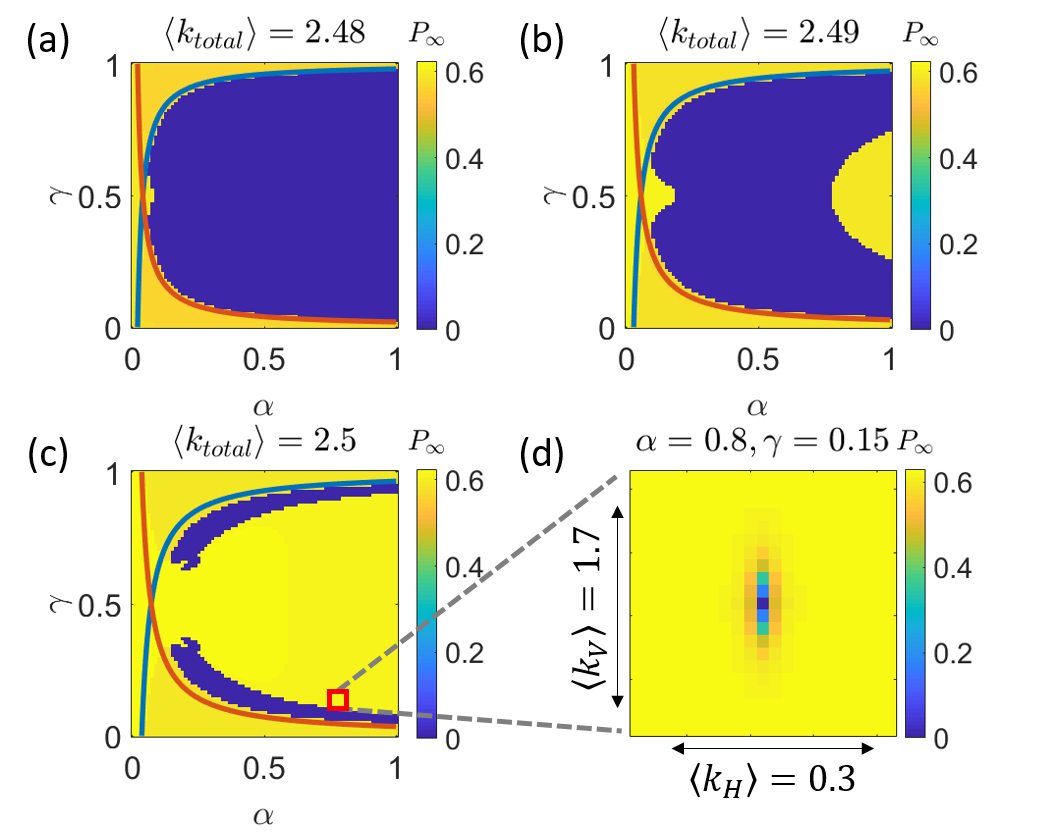}
		\caption{
			\textbf{Robustness after removing one community.}
			\textbf{(a)-(c)} Phase diagrams of the size of the functioning component, $P_\infty$, at steady state after initial removal of one community from a system of 441 communities (i.e. $m=21$). The color of each pixel represents the relative size of the system’s functioning component. 
			In all phase diagrams we show the analytic solutions for $\ktot - \langle k_V \rangle/2 = k_c $ (blue curve) and $\ktot - \langle k_H \rangle/2 = k_c $ (orange curve), where $k_c = 2.4554$ \cite{buldyrev-nature2010}.
			\textbf{(d)} The network at steady state for the case of $\alpha = 0.8$, $\gamma = 0.15$ and $\ktot = 2.5$.
			Each pixel represents a community, and the color of the pixel represents the relative size of the functioning component of the community.
			Since $\gamma<1/2$, the network is anisotropic with more vertical-links than horizontal-links (see Eq. (\ref{eq:gamma})). The initial damage of removing one community causes cascading failures which spread more in the vertical axis than in the horizontal axis, i.e., along the direction of the higher degree.
		}
		\label{fig:Phase_diagrams_small_degree}
	\end{figure}
	
	\begin{figure}
		\centering
		\includegraphics[width=1\linewidth]{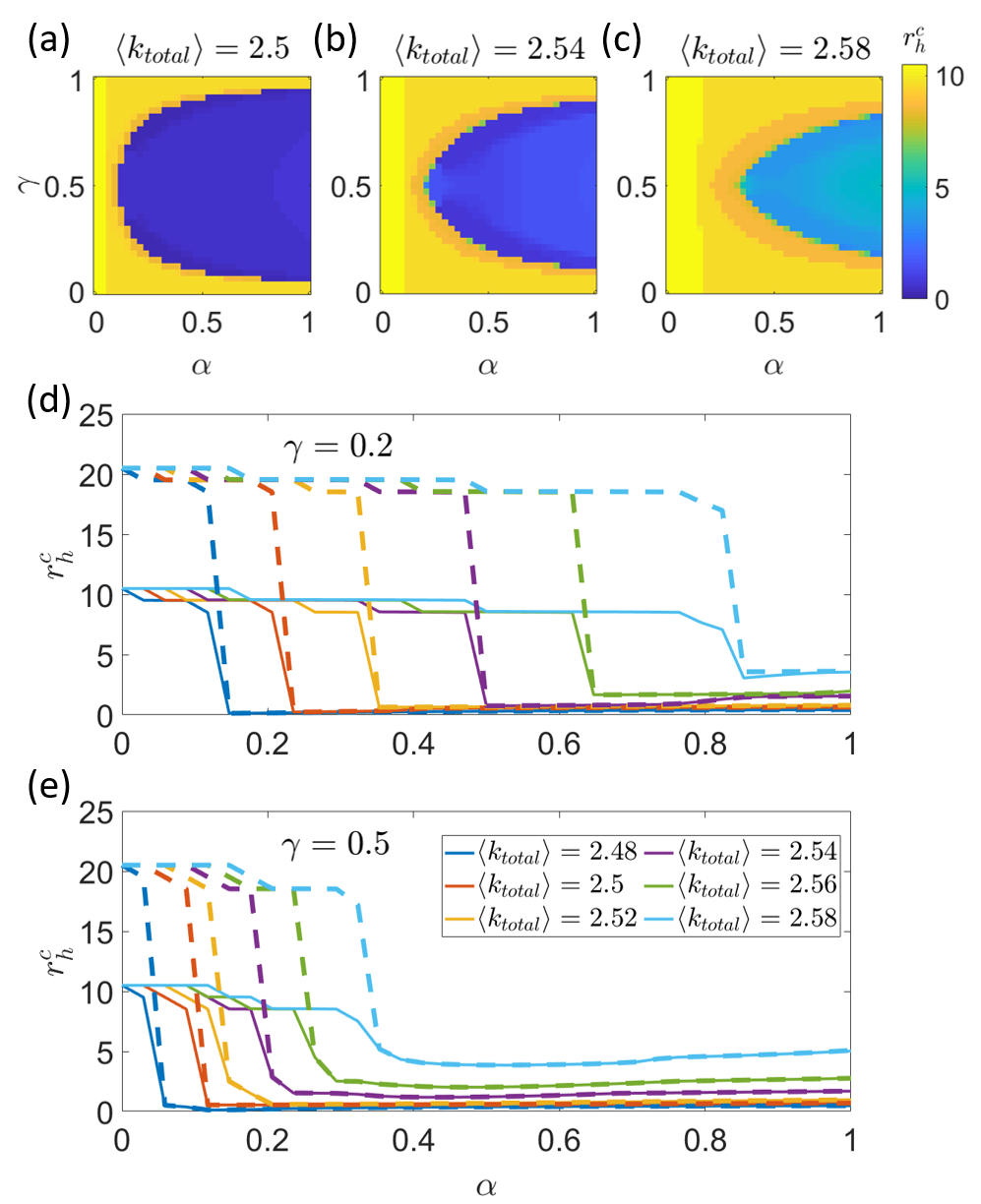}
		\caption{
			\textbf{The critical radius as a function of $\alpha$ and $\gamma$.}
			\textbf{(a)-(c)} Phase diagrams for the critical circular hole radius (in units of a community size), $r_h^c$, for different average total degree $\ktot$. 
			In these simulations we set the system size to be $m=21$.
			\textbf{(d)-(e)} The critical radius, $r_h^c$, for two values of $m$.
			The dashed lines show the $r_h^c$ for the case of $m=41$ and the continuous lines are for the case of $m=21$.
            We find a metastable region, where a finite-size localized attack larger than $r_h^c$, which \textit{does not} depend on the system size $m$, causes cascading failures and system collapse. 
			The metastable region reduces as $\ktot$ increases, and for a fixed $\ktot$, there exists a critical value of $\alpha$ above which the damage of finite radius causes the collapse of the network. The critical $\alpha$ increases both by $\ktot$ and anisotropy.
		}
		\label{fig:Phase_diagrams_large_degree}
	\end{figure}
	
	\begin{figure}
		\centering
		\includegraphics[width=1\linewidth]{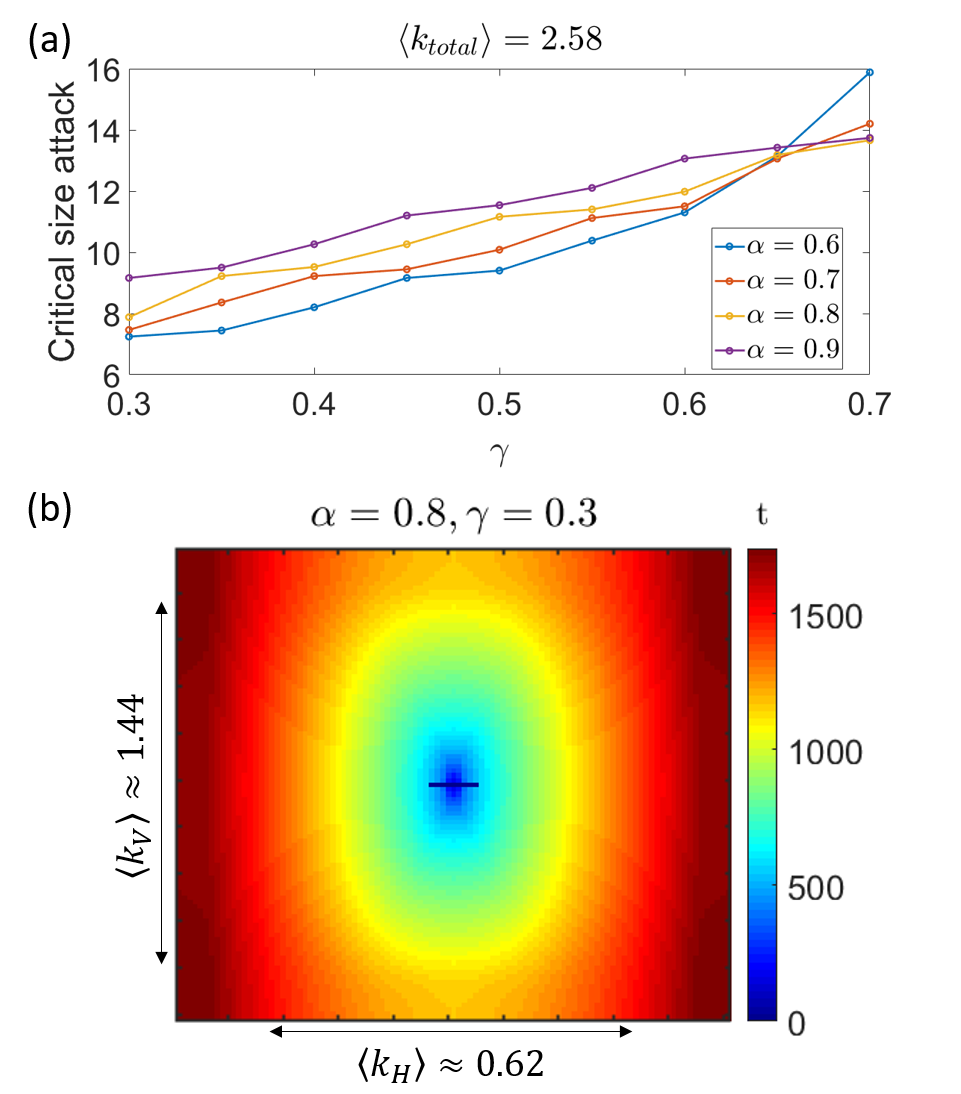}
		\caption{
		\textbf{Critical strip size.}
        \textbf{(a)} The critical size (in units of a community size) for an attack in a form of a strip with a thickness of $1$ community. 
        For $\gamma<1/2$ the strip is orthogonal to the direction of the higher degree interlinks and for $\gamma>1/2$ the strip is parallel to that direction.
        This critical size increases with $\gamma$ and thus it can be concluded that for an efficient attack on an anisotropic network it is more advisable to remove a strip perpendicular to the direction of the higher degree interlinks.
        For these simulations $m=21$ and $\ktot=2.58$.
        \textbf{(b)} Demonstration of the evolution of cascading failures after an attack of length $10$, which is larger than the critical size.
        Each pixel represents a community and the color represents the time $t$ when the community completely collapsed. 
        For this simulation $m=100$, $\ktot=2.58$, $\alpha=0.8$ and $\gamma=0.3$.
		}
		\label{fig:critical_strip}
	\end{figure}

	\begin{figure*}[ht]
		\centering
		\includegraphics[width=1\linewidth]{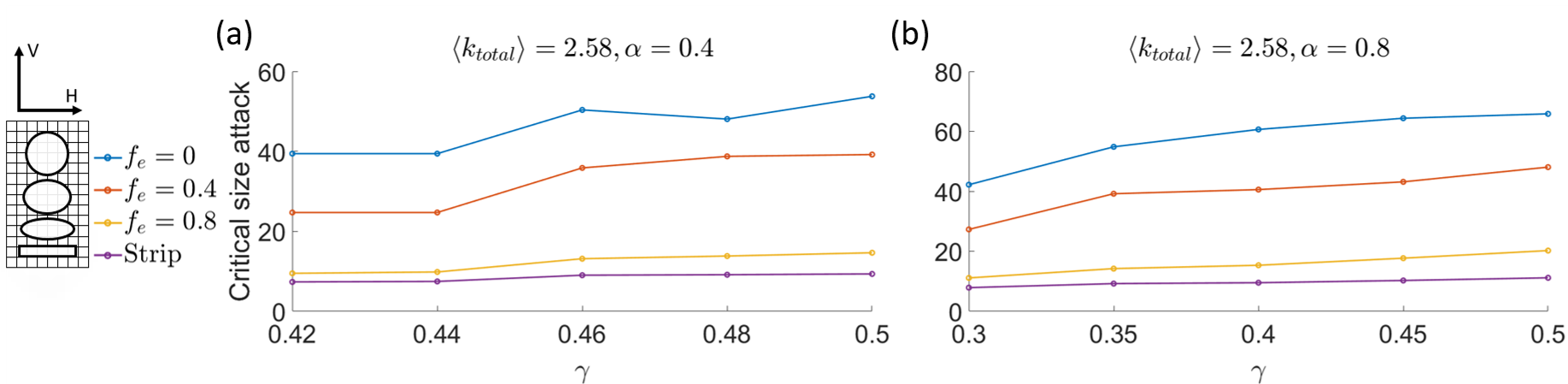}
		\caption{
			\textbf{Comparing attacks of different shapes.}
			Critical size for attacks in a form of an ellipse of different flattening $f_e$ with a major axis orthogonal to the direction of the higher inter degree. 
			The critical size of the attack is measured as the area of the originally deleted communities. 
			Since we remove all the nodes within an ellipse, the communities on the ellipse border are partially removed.
			In addition, we compare the different results to the results of removal of a strip of thickness 1 measured in community sizes.
			An illustration of the attacks for $\kh\leq\kv$ (i.e., $\gamma\leq 0.5$) is presented on the left, where the lattice of communities is at the background.
            We calculate the critical attacks for different values of $\gamma$, for two values of heterogeneity: \textbf{(a)} $\alpha = 0.4$ and \textbf{(b)} $\alpha = 0.8$. In detail, we select different ranges of $\gamma$ for each $\alpha$ in order to be in the metastable region, where the size of the attack does not depend on $m$ (see Appendix Fig. \ref{fig:diff_attacks_m2141}).
            We find that attacks in the form of a strip are more effective than in the form of an ellipse or a circle.
            Surprisingly, those attacks are more effective even in isotropic systems ($\gamma=0.5$).
            For all simulations $m=21$ and $\ktot=2.58$.
		}
		\label{fig:diff_attacks}
	\end{figure*}
    
    Fig. \ref{fig:Phase_diagrams_small_degree} shows that for a given $\ktot$, the robustness of the network to a removal of one community is highly dependent on both parameters $\alpha$ and $\gamma$ representing the heterogeneity and anisotropy, respectively. 
    In particular, in the phase diagrams for $\ktot = 2.48,2.49,2.5$ (Figs. \ref{fig:Phase_diagrams_small_degree}(a)-(c)), the steady state of the system can be in one of two extreme states: stable (in yellow) - in which the system remains functional, and unstable (in blue) - in which the entire system collapses.
    It is important to note that the regions of the stable and unstable states do not depend on the system size $m$ (see also Fig. \ref{fig:Phase_diagrams_small_degree_m2141} in the Appendix).
    In addition, the transition from stable to unstable is non-monotonic with the anisotropy of the network. 
    For instance, for $\ktot = 2.5$ and $\alpha = 0.5$ (see Fig.\ref{fig:Phase_diagrams_small_degree}(c)), the unstable region is only in a specific narrow region of $\gamma$ that is located between the extreme cases of $\gamma = 0.5$ (isotropic network) and $\gamma = 0$ or $\gamma = 1$ (highly anistortopic network).
    
    The blue and orange curves added to the phase diagrams (Figs. \ref{fig:Phase_diagrams_small_degree}(a)-(c)) show that the unstable region is contained within a specific area which is between the two curves and to the right of their intersection. 
    This area is defined by the following constraints: $\ktot - \langle k_H \rangle/2 < k_c$ and $\ktot - \langle k_V \rangle/2 < k_c$, where $k_c \approx 2.4554$ is the critical average degree below which a single ER multiplex collapses without any initial damage \cite{buldyrev-nature2010}.
    These constraints describe a case when removing a community causes its neighboring communities to reduce their average total degree below $k_c$.
    In this mentioned area, there is also a stable region because the communities are not isolated and therefore a community with a total degree smaller than $k_c$ can be sustained by the communities next to it.
    In Fig. \ref{fig:Phase_diagrams_small_degree}(d) we show the behavior of cascading failures in this stable region.
    We show that the initial damage of removing one community spreads further along the direction of the \textit{higher} inter-degree.
    
    Next, we determine the robustness of the system to various forms of isotropic and anisotropic localized attacks. 
    We find that our system is metastable for a broad range of parameters, meaning that for a localized attack of a given shape there is a critical size above which the induced cascade of failures propagates through the whole system leading to its collapse.
    We observe that the size of the critical attack, which does not depend on the number of communities $m$, is significantly different for various shapes of attacks.

    In Fig. \ref{fig:Phase_diagrams_large_degree} we analyze the case of isotropic localized attacks in a form of a circle with a radius $r_h$. 
    We show that for a given $\ktot$ and $\gamma$, there is a critical $\alpha_c$ which has the following properties. For heterogeneity of $\alpha < \alpha_c$ the critical size $r_h^c$ is of size $\approx m/2$ (system size), and for $\alpha > \alpha_c$, $r_h^c$ does not depend on the system size and may contain more than one community. 
    In Fig. \ref{fig:critical_strip}(a) we present the critical attack size for attacks in a form of a strip, i.e, removing communities connected in a row.
    We find that the attack is more efficient when the removed strip is orthogonal to the direction of the higher degree so it destroys more interlinks per one removed community.
    In addition, we find that increasing the anisotropy of the network causes the critical length of the stripe attack to decrease. 
    Note that we have chosen to focus on the range of anisotropy $0.3<\gamma<0.7$ because outside this range the system is always stable, i.e. the critical attack size is in the size of the system (see Fig. \ref{fig:Phase_diagrams_large_degree}).
    Note also that the small fluctuations are caused by the fact that the network is not continuous but consists of discrete communities, i.e., there are jumps when the damage crosses neighboring communities.
    
    In addition, in Fig. \ref{fig:critical_strip}(b), we analyze and demonstrate the evolution of cascading failures after a strip attack whose length is above the critical size. 
    For these simulations, we solved the general iterative equations from \cite{vaknin-prr2020}, which represent the stages of the cascading failures, for our specific anisotropic model (see Eq. (\ref{eq:lattice_fj_gj}) in the Appendix).
    We find that the damage propagates first in the axis with the higher inter degree and only after the damage reaches a certain distance in this axis, it begins to propagate also in the axis with the lower inter degree.
    From this we conclude that the cascading along the axis is affected more by the size of the front of the attack and less by the depth of the attack.
    
    Finally, we analyze the robustness of our anisotropic model with respect to localized attacks in a form of an ellipse with different flattening $f_e$, defined as $f_e = (a-b)/a$ where $a$ and $b$ are the semi-major and semi-minor axes, respectively. 
	The case $f_e=0$ corresponds to our earlier circular attack while $f_e \rightarrow 1$ represents a strip-like attack.
	Thus, the general ellipse case covers a wide range of anisotropic attacks between a strip attack and a circle attack. 
    In Fig. \ref{fig:diff_attacks} we show that the ellipse attack is more efficient for ellipses with higher flattening $f_e$, and that a strip attack is more efficient than any ellipse attack.
    We therefore conclude that for a given area of the attack, it is more efficient that the shape of the attack will be as anisotropic as possible.
    
 	\section{Discussion}
    
    Many spatial networks are influenced by geographical features that can lead to an anisotropic structure in which the amount of links differs in different directions.
    In this work, we have introduced a new realistic model, which considers, for the first time, the aspect of network anisotropy.
    Using tools from percolation theory, we systematically study anisotropic systems and analyze their robustness to various isotropic and anisotropic localized attacks.
    We determine how the robustness of our network model depends on its heterogeneity and anisotropy. 
    We also show that anisotropic attacks reveal significantly increased vulnerability compared to the considered earlier, simple circle-shaped attacks.
    Specifically, we find that a localized attack causes larger damage if the area of the attack is more elongated in the direction of the smaller inter degree, i.e. when the attack cuts more links that are in the direction of the higher inter degree.
    We find that even in isotropic networks the anisotropic localized attacks are more efficient than isotropic attacks, see the case $\gamma = 0.5$ in Fig. \ref{fig:diff_attacks}.
    In our future work, we will determine how the velocity of the cascading failures in the different axes depends on the parameters of the anisotropic network, and will try to answer the question of whether this velocity exhibits a scaling behavior.
    
 	\section*{ACKNOWLEDGMENTS}

    We thank the Israel Science Foundation, the Binational Israel-China Science Foundation (Grant No. 3132/19), the BIU Center for Research in Applied Cryptography and Cyber Security, NSF-BSF (Grant No. 2019740), the EU H2020 project RISE (Project No. 821115), the EU H2020 DIT4TRAM, and DTRA (Grant No. HDTRA-1-19-1-0016) for financial support.
    L.A.B acknowledges UNMdP (EXA 956/20), for financial support. 
    D.V. thanks the PBC of the Council for Higher Education of Israel for the Fellowship Grant and Lucas D. Valdez for valuable discussions.

	\FloatBarrier

	\section{APPENDIX}
    
    \subsection{Comparison between simulation results and theory}
    
\begin{figure}[ht!]
		\centering
		\includegraphics[width=1\linewidth]{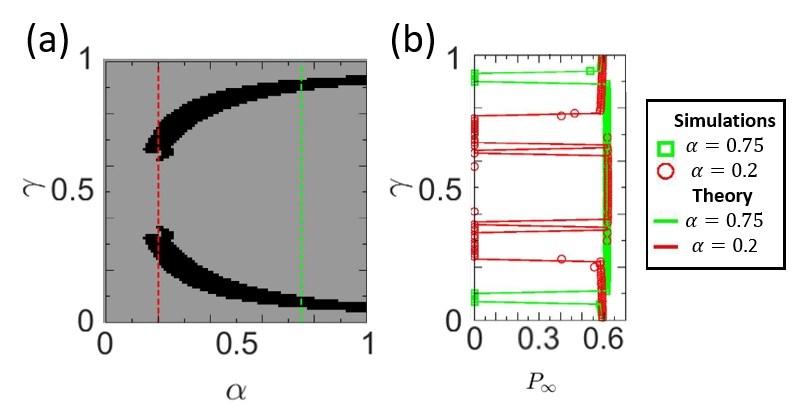}
		\caption{
			\textbf{Simulations and analytical results.}
			\textbf{(a)} Phase diagram of $P_\infty$ after initial attack of removing one community, for different values of $\alpha$ and $\gamma$ (as in Fig. \ref{fig:Phase_diagrams_small_degree}). 
			\textbf{(b)} $P_\infty$ as a function of $\gamma$ for two values of $\alpha$, which are highlighted with dashed vertical lines in \textbf{(a)}.
			The lines represent the theory of Eq. (\ref{eq:Pinf}), and symbols are simulations averaged over 100 realizations on networks with community size $\zeta = 201$.
			For both figures, $m = 21$ and $\ktot = 2.5$.
		}
		\label{fig:Phase_diagrams_simulations}
	\end{figure}

    In Fig. \ref{fig:Phase_diagrams_simulations}, we compare between the theory Eq. (\ref{eq:Pinf}) and numerical simulations for $P_\infty$ after initial removal of one community and observe excellent agreement between them. 

    \subsection{The cascading failures equations}
    
    Here we present, for the convenience, the general equations from our previous article \cite{vaknin-prr2020}, and apply them to the anisotropic model of this manuscript.
    For a multiplex model with two layers $A$ and $B$, the vector equations of the cascading failures starting from $t=0$ are:
    \begin{equation}
    \label{eq:cascade}
    \begin{gathered}
    \vec{f}_A(2t)=\vec{\Phi}_A[\vec{f}_A(2t),\vec{p}(2t)]\\
    \vec{g}_A(2t)=\vec{\Psi}_A[\vec{f}_A(2t),\vec{p}(2t)]\\
    \vec{p}(2t+1)=\vec{\Psi}_A[\vec{f}_A(2t),\vec{p}(0)]\\
    \vec{f}_B(2t+1)=\vec{\Phi}_B[\vec{f}_B(2t+1),\vec{p}(2t+1)]\\
    \vec{g}_B(2t+1)=\vec{\Psi}_B[\vec{f}_B(2t+1),\vec{p}(2t+1)]\\
    \vec{p}(2t+2)=\vec{\Psi}_B[\vec{f}_B(2t+1),\vec{p}(0)],
    \end{gathered}
    \end{equation}
    where the parameters are defined as follows, (1) $f_{i,j}$ is the probability that a randomly selected link, that passes from a node in community $i$ to a node in community $j$, does not lead to the giant component (GC), (2) $g_i$ is the fraction of nodes in community $i$ which belong to the GC, and (3) $\vec{p}(t)$ is the fraction of survived nodes at stage $t$ of the cascade of failures.
    For our anisotropic multiplex model, in which the degree distributions are the same in both layers, and in the limit of infinitely large ER communities, the functions $\vec{\Phi}$ and $\vec{\Psi}$ are obtained by 
    \begin{equation}
    \begin{gathered}
    f_j = \vec{\Phi}_j(\vec{f},\vec{p}) \equiv (1-p_j) + p_j \cdot  e^{\sum_{i} \mkkij \cdot (f_i -1)}\\
    \label{eq:lattice_fj_gj}
    g_j = \vec{\Psi}_j(\vec{f},\vec{p})\equiv p_j(1-e^{\sum_{i} \mkkij \cdot (f_i -1)}),
    \end{gathered}
    \end{equation}
    where $f_j\equiv f_{i,j}$ and $\langle k_{i,j}\rangle$ is defined in Eq.(\ref{eq:kij}) in the main text.
    
    \subsection{The analytical result for different system size}
    
\begin{figure}[ht!]
		\centering
		\includegraphics[width=1\linewidth]{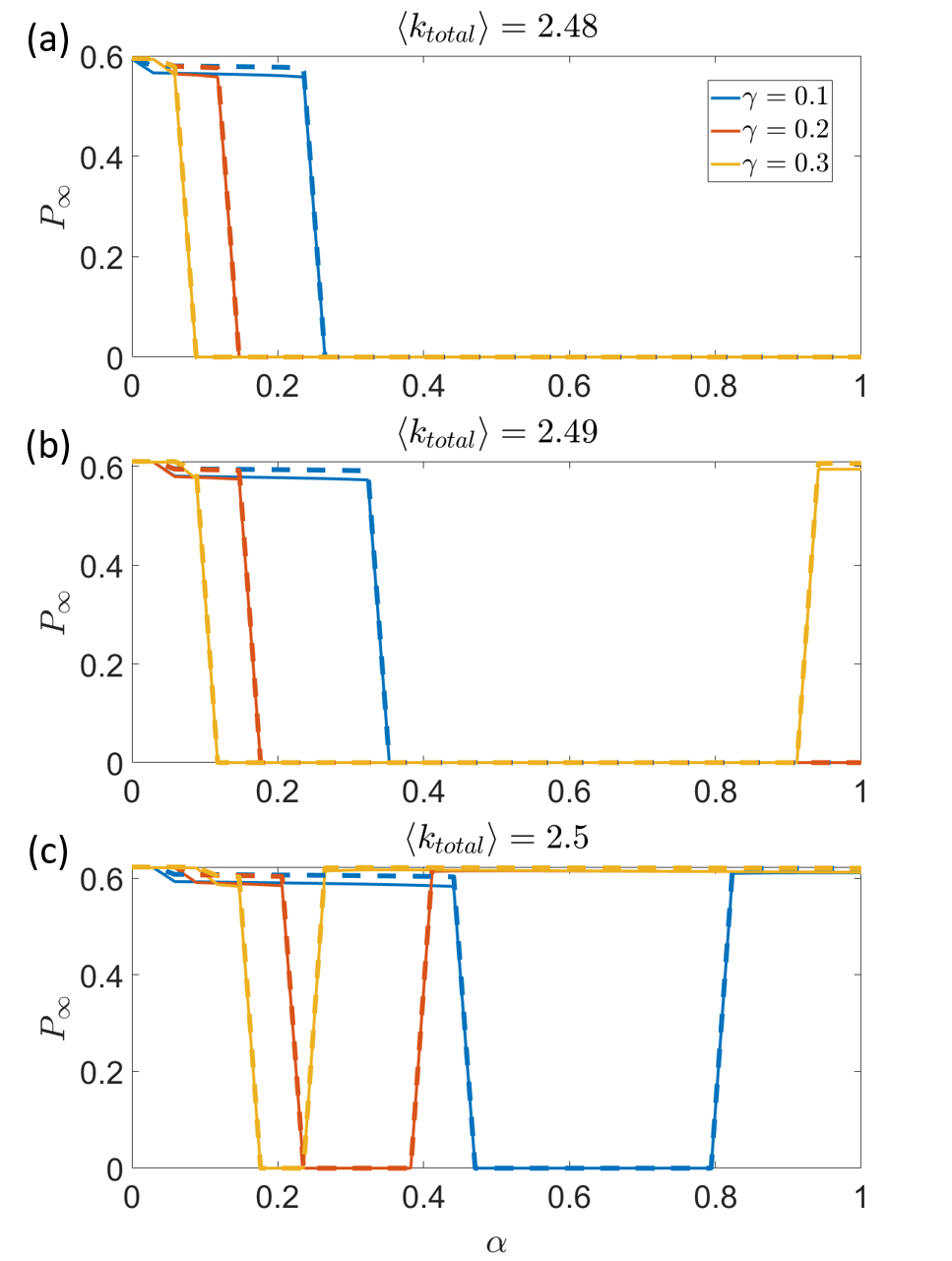}
		\caption{
			\textbf{Robustness after removing one community for two values of $m$.}
			\textbf{(a)-(c)} The functioning component, $P_\infty$, at steady state after initial removal of one community for different average total degree $\ktot$. The dashed lines show the $P_\infty$ for the case of $m=41$ and the continuous line is for the case of $m=21$.
			The regions in which the system is stable ($P_\infty \approx  0.6$) and unstable ($P_\infty = 0$) do not depend on the size of the system.
		}
		\label{fig:Phase_diagrams_small_degree_m2141}
	\end{figure}

\begin{figure}[ht!]
		\centering
		\includegraphics[width=1\linewidth]{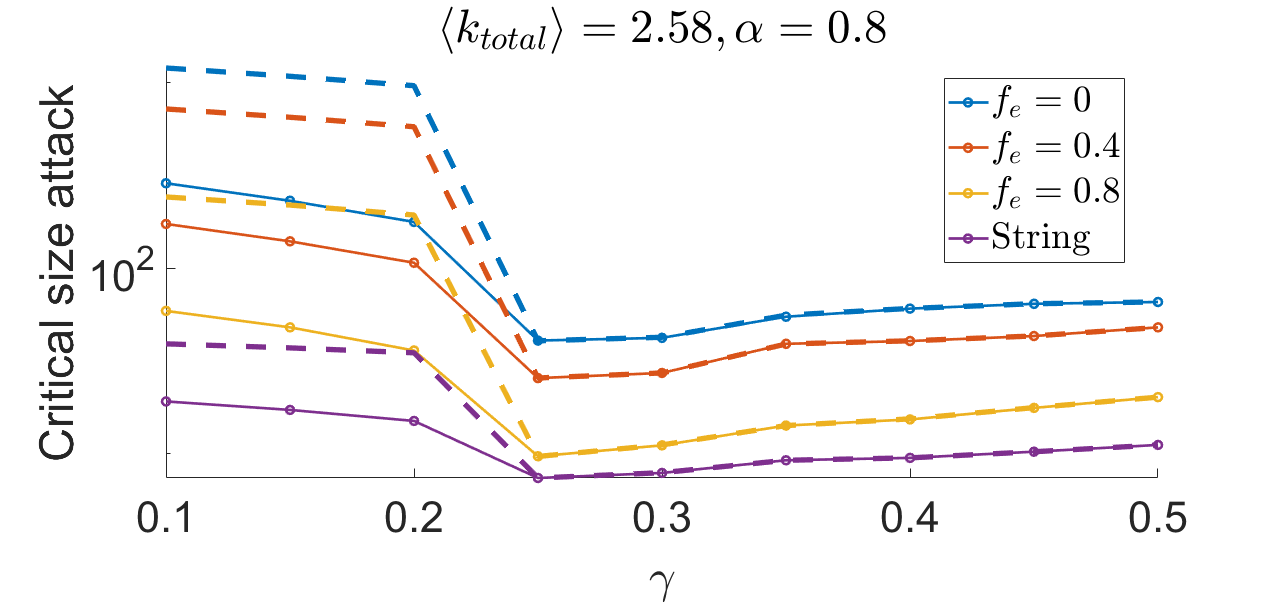}
		\caption{
		\textbf{Comparing attacks of different shapes for two values of $m$.}
		The critical size attack for attacks in an ellipse form and for removing a strip, on a log-linear graph, for a wide range of $\gamma$ for $\ktot=2.58$ and $\alpha = 0.8$ (as in Fig. \ref{fig:diff_attacks}(b)). 
		The continuous lines represent $m = 21$ and the dashed lines represent $m = 41$.
		For $\gamma < 0.25$ the system size $m$ affects the size of the critical attack, which is the whole system.
		In contrast, for $\gamma \geq 0.25$ the critical size does not depend on the system size $m$.
		}
		\label{fig:diff_attacks_m2141}
	\end{figure}

    Figures \ref{fig:Phase_diagrams_small_degree_m2141} and \ref{fig:diff_attacks_m2141} show that there is no effect of the system size $m$ on our analytical results.
    
	\FloatBarrier
	\bibliographystyle{naturemag_4etal}
	\bibliography{mybib}

\begin{thebibliography}{10}
\expandafter\ifx\csname url\endcsname\relax
  \def\url#1{\texttt{#1}}\fi
\expandafter\ifx\csname urlprefix\endcsname\relax\def\urlprefix{URL }\fi
\providecommand{\bibinfo}[2]{#2}
\providecommand{\eprint}[2][]{\url{#2}}

\bibitem{girvan-pnas2002}
\bibinfo{author}{Girvan, M.} \& \bibinfo{author}{Newman, M. E.~J.}
\newblock \bibinfo{title}{Community structure in social and biological
  networks}.
\newblock \emph{\bibinfo{journal}{Proceedings of the National Academy of
  Sciences}} \textbf{\bibinfo{volume}{99}}, \bibinfo{pages}{7821--7826}
  (\bibinfo{year}{2002}).

\bibitem{guimera-pnas2005}
\bibinfo{author}{Guimera, R.} \emph{et~al.}
\newblock \bibinfo{title}{The worldwide air transportation network: Anomalous
  centrality, community structure, and cities global roles}.
\newblock \emph{\bibinfo{journal}{Proceedings of the National Academy of
  Sciences}} \textbf{\bibinfo{volume}{102}}, \bibinfo{pages}{7794--7799}
  (\bibinfo{year}{2005}).

\bibitem{palla-nature2005}
\bibinfo{author}{Palla, G.} \emph{et~al.}
\newblock \bibinfo{title}{Uncovering the overlapping community structure of
  complex networks in nature and society}.
\newblock \emph{\bibinfo{journal}{Nature}} \textbf{\bibinfo{volume}{435}},
  \bibinfo{pages}{814--818} (\bibinfo{year}{2005}).

\bibitem{kosmidis-epl2008}
\bibinfo{author}{Kosmidis, K.}, \bibinfo{author}{Havlin, S.} \&
  \bibinfo{author}{Bunde, A.}
\newblock \bibinfo{title}{Structural properties of spatially embedded
  networks}.
\newblock \emph{\bibinfo{journal}{EPL (Europhysics Letters)}}
  \textbf{\bibinfo{volume}{82}}, \bibinfo{pages}{48005} (\bibinfo{year}{2008}).

\bibitem{bradde-prl2010}
\bibinfo{author}{Bradde, S.} \emph{et~al.}
\newblock \bibinfo{title}{Critical fluctuations in spatial complex networks}.
\newblock \emph{\bibinfo{journal}{Phys. Rev. Lett.}}
  \textbf{\bibinfo{volume}{104}}, \bibinfo{pages}{218701}
  (\bibinfo{year}{2010}).

\bibitem{mucha-science2010}
\bibinfo{author}{Mucha, P.~J.} \emph{et~al.}
\newblock \bibinfo{title}{Community structure in time-dependent, multiscale,
  and multiplex networks}.
\newblock \emph{\bibinfo{journal}{Science}} \textbf{\bibinfo{volume}{328}},
  \bibinfo{pages}{876--878} (\bibinfo{year}{2010}).

\bibitem{barthelemy-physicsreports2011}
\bibinfo{author}{Barth{\'e}l\'emy, M.}
\newblock \bibinfo{title}{Spatial networks}.
\newblock \emph{\bibinfo{journal}{Physics Reports}}
  \textbf{\bibinfo{volume}{499}}, \bibinfo{pages}{1 -- 101}
  (\bibinfo{year}{2011}).

\bibitem{bashan-naturephysics2013}
\bibinfo{author}{Bashan, A.} \emph{et~al.}
\newblock \bibinfo{title}{{The extreme vulnerability of interdependent
  spatially embedded networks}}.
\newblock \emph{\bibinfo{journal}{Nature Physics}}
  \textbf{\bibinfo{volume}{9}}, \bibinfo{pages}{667--672}
  (\bibinfo{year}{2013}).

\bibitem{du-chaos2014}
\bibinfo{author}{Du, W.-B.} \emph{et~al.}
\newblock \bibinfo{title}{Traffic dynamics on coupled spatial networks}.
\newblock \emph{\bibinfo{journal}{Chaos, Solitons \& Fractals}}
  \textbf{\bibinfo{volume}{68}}, \bibinfo{pages}{72--77}
  (\bibinfo{year}{2014}).

\bibitem{danziger-epl2016}
\bibinfo{author}{Danziger, M.~M.} \emph{et~al.}
\newblock \bibinfo{title}{The effect of spatiality on multiplex networks}.
\newblock \emph{\bibinfo{journal}{EPL (Europhysics Letters)}}
  \textbf{\bibinfo{volume}{115}}, \bibinfo{pages}{36002}
  (\bibinfo{year}{2016}).

\bibitem{rinaldi-ieee2001}
\bibinfo{author}{Rinaldi, S.}, \bibinfo{author}{Peerenboom, J.} \&
  \bibinfo{author}{Kelly, T.}
\newblock \bibinfo{title}{{Identifying, understanding, and analyzing critical
  infrastructure interdependencies}}.
\newblock \emph{\bibinfo{journal}{Control Systems, IEEE}}
  \textbf{\bibinfo{volume}{21}}, \bibinfo{pages}{11--25}
  (\bibinfo{year}{2001}).

\bibitem{chang-bridge2009}
\bibinfo{author}{Chang, S.~E.}
\newblock \bibinfo{title}{Infrastructure resilience to disasters}.
\newblock \emph{\bibinfo{journal}{The Bridge}} \textbf{\bibinfo{volume}{39}},
  \bibinfo{pages}{36--41} (\bibinfo{year}{2009}).

\bibitem{hines-proceedings2010}
\bibinfo{author}{Hines, P.} \emph{et~al.}
\newblock \bibinfo{title}{{The Topological and Electrical Structure of Power
  Grids}}.
\newblock In \emph{\bibinfo{booktitle}{{System Sciences (HICSS), 2010 43rd
  Hawaii International Conference on}}}, \bibinfo{pages}{1--10}
  (\bibinfo{year}{2010}).

\bibitem{buldyrev-nature2010}
\bibinfo{author}{Buldyrev, S.~V.} \emph{et~al.}
\newblock \bibinfo{title}{{Catastrophic cascade of failures in interdependent
  networks}}.
\newblock \emph{\bibinfo{journal}{Nature}} \textbf{\bibinfo{volume}{464}},
  \bibinfo{pages}{1025--1028} (\bibinfo{year}{2010}).

\bibitem{parshani-prl2010}
\bibinfo{author}{Parshani, R.}, \bibinfo{author}{Buldyrev, S.~V.} \&
  \bibinfo{author}{Havlin, S.}
\newblock \bibinfo{title}{{Interdependent Networks: Reducing the Coupling
  Strength Leads to a Change from a First to Second Order Percolation
  Transition}}.
\newblock \emph{\bibinfo{journal}{Phys. Rev. Lett.}}
  \textbf{\bibinfo{volume}{105}}, \bibinfo{pages}{048701}
  (\bibinfo{year}{2010}).

\bibitem{gao-pre2012}
\bibinfo{author}{Gao, J.} \emph{et~al.}
\newblock \bibinfo{title}{Robustness of a network formed by $n$ interdependent
  networks with a one-to-one correspondence of dependent nodes}.
\newblock \emph{\bibinfo{journal}{Phys. Rev. E}} \textbf{\bibinfo{volume}{85}},
  \bibinfo{pages}{066134} (\bibinfo{year}{2012}).

\bibitem{baxter-prl2012}
\bibinfo{author}{Baxter, G.~J.} \emph{et~al.}
\newblock \bibinfo{title}{{Avalanche Collapse of Interdependent Networks}}.
\newblock \emph{\bibinfo{journal}{Phys. Rev. Lett.}}
  \textbf{\bibinfo{volume}{109}}, \bibinfo{pages}{248701}
  (\bibinfo{year}{2012}).

\bibitem{dedomenico-prx2013}
\bibinfo{author}{De~Domenico, M.} \emph{et~al.}
\newblock \bibinfo{title}{Mathematical formulation of multilayer networks}.
\newblock \emph{\bibinfo{journal}{Phys. Rev. X}} \textbf{\bibinfo{volume}{3}},
  \bibinfo{pages}{041022} (\bibinfo{year}{2013}).

\bibitem{son-epl2012}
\bibinfo{author}{Son, S.-W.} \emph{et~al.}
\newblock \bibinfo{title}{{Percolation theory on interdependent networks based
  on epidemic spreading}}.
\newblock \emph{\bibinfo{journal}{EPL (Europhysics Letters)}}
  \textbf{\bibinfo{volume}{97}}, \bibinfo{pages}{16006} (\bibinfo{year}{2012}).

\bibitem{dong-pre2014}
\bibinfo{author}{Zhou, D.} \emph{et~al.}
\newblock \bibinfo{title}{Simultaneous first- and second-order percolation
  transitions in interdependent networks}.
\newblock \emph{\bibinfo{journal}{Phys. Rev. E}} \textbf{\bibinfo{volume}{90}},
  \bibinfo{pages}{012803} (\bibinfo{year}{2014}).

\bibitem{kivela-jcomnets2014}
\bibinfo{author}{Kivel{\"a}, M.} \emph{et~al.}
\newblock \bibinfo{title}{Multilayer networks}.
\newblock \emph{\bibinfo{journal}{Journal of Complex Networks}}
  \textbf{\bibinfo{volume}{2}}, \bibinfo{pages}{203--271}
  (\bibinfo{year}{2014}).

\bibitem{boccaletti-physicsreports2014}
\bibinfo{author}{Boccaletti, S.} \emph{et~al.}
\newblock \bibinfo{title}{The structure and dynamics of multilayer networks}.
\newblock \emph{\bibinfo{journal}{Physics Reports}}
  \textbf{\bibinfo{volume}{544}}, \bibinfo{pages}{1--122}
  (\bibinfo{year}{2014}).

\bibitem{radicchi-naturephysics2015}
\bibinfo{author}{Radicchi, F.}
\newblock \bibinfo{title}{Percolation in real interdependent networks}.
\newblock \emph{\bibinfo{journal}{Nature Physics}}
  \textbf{\bibinfo{volume}{11}}, \bibinfo{pages}{597–602}
  (\bibinfo{year}{2015}).

\bibitem{shekhtman-njp2015}
\bibinfo{author}{Shekhtman, L.~M.}, \bibinfo{author}{Shai, S.} \&
  \bibinfo{author}{Havlin, S.}
\newblock \bibinfo{title}{Resilience of networks formed of interdependent
  modular networks}.
\newblock \emph{\bibinfo{journal}{New Journal of Physics}}
  \textbf{\bibinfo{volume}{17}}, \bibinfo{pages}{123007}
  (\bibinfo{year}{2015}).

\bibitem{bianconi-oxfordjournals2018}
\bibinfo{author}{Bianconi, G.}
\newblock \emph{\bibinfo{title}{Multilayer Networks}}
  (\bibinfo{publisher}{Oxford University Press}, \bibinfo{year}{2018}).

\bibitem{motter-pre2002}
\bibinfo{author}{Motter, A.~E.} \& \bibinfo{author}{Lai, Y.-C.}
\newblock \bibinfo{title}{Cascade-based attacks on complex networks}.
\newblock \emph{\bibinfo{journal}{Phys. Rev. E}} \textbf{\bibinfo{volume}{66}},
  \bibinfo{pages}{065102} (\bibinfo{year}{2002}).

\bibitem{wei-prl2012}
\bibinfo{author}{Li, W.} \emph{et~al.}
\newblock \bibinfo{title}{{Cascading Failures in Interdependent Lattice
  Networks: The Critical Role of the Length of Dependency Links}}.
\newblock \emph{\bibinfo{journal}{Phys. Rev. Lett.}}
  \textbf{\bibinfo{volume}{108}}, \bibinfo{pages}{228702}
  (\bibinfo{year}{2012}).

\bibitem{gao-naturephysics2012}
\bibinfo{author}{Gao, J.} \emph{et~al.}
\newblock \bibinfo{title}{{Networks formed from interdependent networks}}.
\newblock \emph{\bibinfo{journal}{Nature Physics}}
  \textbf{\bibinfo{volume}{8}}, \bibinfo{pages}{40--48} (\bibinfo{year}{2012}).

\bibitem{reis-naturephysics2014}
\bibinfo{author}{Reis, S. D.~S.} \emph{et~al.}
\newblock \bibinfo{title}{Avoiding catastrophic failure in correlated networks
  of networks}.
\newblock \emph{\bibinfo{journal}{Nature Physics}}
  \textbf{\bibinfo{volume}{10}}, \bibinfo{pages}{762–767}
  (\bibinfo{year}{2014}).

\bibitem{yuan-pre2015}
\bibinfo{author}{Yuan, X.} \emph{et~al.}
\newblock \bibinfo{title}{How breadth of degree distribution influences network
  robustness: comparing localized and random attacks}.
\newblock \emph{\bibinfo{journal}{Physical Review E}}
  \textbf{\bibinfo{volume}{92}}, \bibinfo{pages}{032122}
  (\bibinfo{year}{2015}).

\bibitem{vaknin-njp2017}
\bibinfo{author}{Vaknin, D.}, \bibinfo{author}{Danziger, M.~M.} \&
  \bibinfo{author}{Havlin, S.}
\newblock \bibinfo{title}{Spreading of localized attacks in spatial multiplex
  networks}.
\newblock \emph{\bibinfo{journal}{New Journal of Physics}}
  \textbf{\bibinfo{volume}{19}}, \bibinfo{pages}{073037}
  (\bibinfo{year}{2017}).

\bibitem{spiewak-netwsci2018}
\bibinfo{author}{Spiewak, R.} \emph{et~al.}
\newblock \bibinfo{title}{A study of cascading failures in real and synthetic
  power grid topologies.}
\newblock \emph{\bibinfo{journal}{Netw. Sci.}} \textbf{\bibinfo{volume}{6}},
  \bibinfo{pages}{448--468} (\bibinfo{year}{2018}).

\bibitem{berezin-scireports2015}
\bibinfo{author}{Berezin, Y.} \emph{et~al.}
\newblock \bibinfo{title}{{Localized attacks on spatially embedded networks
  with dependencies}}.
\newblock \emph{\bibinfo{journal}{Scientific Reports}}
  \textbf{\bibinfo{volume}{5}} (\bibinfo{year}{2015}).

\bibitem{shao-njp2015}
\bibinfo{author}{Shao, S.} \emph{et~al.}
\newblock \bibinfo{title}{Percolation of localized attack on complex networks}.
\newblock \emph{\bibinfo{journal}{New Journal of Physics}}
  \textbf{\bibinfo{volume}{17}}, \bibinfo{pages}{023049}
  (\bibinfo{year}{2015}).

\bibitem{yang-science2017}
\bibinfo{author}{Yang, Y.}, \bibinfo{author}{Nishikawa, T.} \&
  \bibinfo{author}{Motter, A.~E.}
\newblock \bibinfo{title}{Small vulnerable sets determine large network
  cascades in power grids}.
\newblock \emph{\bibinfo{journal}{Science}} \textbf{\bibinfo{volume}{358}}
  (\bibinfo{year}{2017}).

\bibitem{vaknin-prr2020}
\bibinfo{author}{Vaknin, D.} \emph{et~al.}
\newblock \bibinfo{title}{Spreading of localized attacks on spatial multiplex
  networks with a community structure}.
\newblock \emph{\bibinfo{journal}{Phys. Rev. Research}}
  \textbf{\bibinfo{volume}{2}}, \bibinfo{pages}{043005} (\bibinfo{year}{2020}).

\end{thebibliography}
	
\end{document}